\documentclass[11pt]{article}

\usepackage{graphicx}
\graphicspath{{Figures/}}

\usepackage[english]{babel}
\usepackage[utf8]{inputenc}
\usepackage{csquotes,amsmath}
\usepackage[letterpaper,top=1in,bottom=1in,left=1in,right=1in,marginparwidth=1in]{geometry}

\usepackage{setspace}
\allowdisplaybreaks

\usepackage{algorithm}
\usepackage{algpseudocode}

\usepackage[citestyle=authoryear-comp, style=authoryear, natbib=true,
maxbibnames=99, giveninits=true,
doi=false,isbn=false,url=false,dashed=false,eprint=false,
uniquename=init,mincitenames=1,maxcitenames=2,
uniquelist=false]{biblatex}
\DeclareNameAlias{author}{last-first}
\bibliography{refs.bib}

\usepackage{amsmath,amsthm,amssymb,bm,bbm}
\usepackage{siunitx}
\usepackage{graphicx}
\usepackage[colorlinks=true, allcolors=blue]{hyperref}

\usepackage{xcolor}

\usepackage{authblk}

\title{Distributional regression for seasonal data: an application to river flows}

\author{Samuel Perreault}
\author{Silvana M. Pesenti}
\author{Daniyal Shahzad}
\affil{Department of Statistical Sciences, University of Toronto, ON, Canada}

\begin{document}
\maketitle
\begin{abstract}
Risk assessment in casualty insurance, such as flood risk, traditionally relies on extreme-value methods that emphasizes rare events. These approaches are well-suited for characterizing tail risk, but do not capture the broader dynamics of environmental variables such as moderate or frequent loss events.  To complement these methods, we propose a modelling framework for estimating the full (daily) distribution of environmental variables as a function of time, that is a distributional version of typical climatological summary statistics, thereby incorporating both seasonal variation and gradual long-term changes. Aside from the time trend, to capture seasonal variation our approach simultaneously estimates the distribution for each instant of the seasonal cycle, without explicitly modelling the temporal dependence present in the data. To do so, we adopt a framework inspired by GAMLSS (Generalized Additive Models for Location, Scale, and Shape), where the parameters of the distribution vary over the seasonal cycle as a function of explanatory variables depending only on the time of year, and not on the past values of the process under study. Ignoring the temporal dependence in the seasonal variation greatly simplifies the modelling but poses inference challenges that we clarify and overcome. 

We apply our framework to daily river flow data from three hydrometric stations along the Fraser River in British Columbia, Canada, and analyse the flood of the Fraser River in early winter of 2021.
\end{abstract}

\textbf{Keywords}: GAMLSS, Climatology, Misspecified model, Environmental risk, Flooding\\

\section{Introduction}
In casualty insurance, risk at a given location can broadly be viewed as the combination of two components: the probability of an event and the exposure, typically measured by the number of people or value of assets in the region. While exposure is often well-documented, understanding how the probability component evolves throughout the year remains a critical challenge—particularly in the context of environmental hazards such as floods. Traditionally, flood risk is summarized using return periods (average time between events of a certain magnitude), which are computed based on annual or seasonal maxima, or using peaks-over-threshold methods \citep{Hamed/Rao:2019,Pan/al:2022}. We also refer to \cite{albrecher2025} who study the spatial and long-term temporal dependence of food risk. These extreme risk approaches are well-established and particularly effective for characterizing tail risk; however, they provide only a partial view of the underlying phenomenon. In particular, they may not fully capture the range of conditions leading to moderate or frequent events, which also have implications for accumulated damages and operational risk. Modelling the full univariate distribution of river flow can serve as a valuable complement to classical extreme-value techniques. In this framework, seasonality and time trend are equally important. In particular, incorporating seasonality in a continuous fashion allows for a more refined understanding of how risk varies throughout the year, which is not possible when only studying seasonal maxima. Allowing for the inclusion of temporal trends makes it possible to account, at least on relatively short time horizons, for evolving patterns potentially associated with climate change. These considerations are highly relevant in insurance and risk management, where understanding the full structure of environmental variability is essential for pricing, underwriting, and capital allocation. 

To motivate our work, we consider daily average river flow data (in \SI{}{m^3/s}) from three hydrometric stations along the Fraser River in British Columbia, Canada. These stations provide long and uninterrupted records, ranging from a few decades to over a century of observations. One of them is located in the Fraser Valley, a region that has experienced major flooding events, including the notable November 2021 flood. The Fraser River is the longest river in British Columbia and exhibits interesting features. It  originates in the Canadian Rocky Mountains, which are heavily snow covered in winter, then passes through the Fraser Canyon to the Fraser Valley, a lowland valley, and then empties into the Strait of Georgia close to Vancouver. The Fraser River is known for flooding, in particular, the Fraser Valley and Vancouver, which due to the proximity to the city results in large losses.

The primary aim of this paper is to introduce a flexible seasonal modelling framework motivated by the need to analyse long-term environmental data with evolving patterns. Building on the methodology proposed by \citet{Perreault/Pesenti/Reid:2025+}, we make two main contributions. First, we extend their purely seasonal model to accommodate temporal trends and interactions between seasonality and long-term variation. This generalization enables the modelling of more complex nonstationary behaviours often observed in environmental processes. Second, we explore the use of the (extended) generalized gamma distribution for modelling river flow data. We find that it offers a highly flexible and accurate fit for this type of data.

\citet{Perreault/Pesenti/Reid:2025+} propose a parametric approach for modelling how the distribution of a seasonal process varies throughout the year. By treating distributional parameters as smooth functions of the day of the year, their method provides a continuous, climatology-like summary of seasonal behaviour that is easily interpretable. It can be understood as a form of distributional regression \citep[see, e.g.,][]{Klein:2024}, and more specifically as a generalized additive model for location, scale, and shape \citep[GAMLSS][]{Rigby/Stasinopoulos:2005}, a modelling framework that is well adopted in hydrology \citep[see, e.g.,][for GAMLSS-based flood frequency analyses]{Villarini/al:2009, Machado/al:2015}. Since temporal dependence is not the target of inference and a relatively complex phenomena, it is purposefully left unspecified, that is we treat the data as independent. This greatly simplifies the model structure but requires appropriate corrections during inference, where we draw from the literature on misspecified models, to account for residual serial correlation. While the focus is on modelling seasonal variation independently of temporal dependence, the proposed methodology can serve as the first step in a more general modelling pipeline, where the resulting fitted distributions are used to construct copula pseudo-observations (via the probability integral transform), from which temporal dependence can then be modelled separately.

A key assumption in the original formulation of \cite{Perreault/Pesenti/Reid:2025+} is that the process is cyclostationary, meaning that the full distribution of the variable repeats itself periodically. Cyclostationary processes have been studied since at least the 1960s \citep{Gladyshev:1963, Brelsford/Jones:1967, Hurd:1970}, and remained an active area of research since then, particularly in signal processing \citep[see, e.g.,][]{Gardner/Napolitano/Paura:2006, Napolitano:2016a, Napolitano:2016b, Napolitano:2020}. While this is a natural and appealing framework for modelling seasonal phenomena, it can be too restrictive in practice, especially when long-term changes or gradual shifts in seasonal behaviour are present. As noted by \citet{Perreault/Pesenti/Reid:2025+}, however, their modelling approach is naturally extensible to accommodate more realistic forms of nonstationarity. In this paper, we do so by introducing slow, non-periodic temporal variation, allowing the seasonal profile to evolve gradually over time.

Our second objective, which is more applied in nature, is to explore the use of the generalized gamma distribution for modelling river flows. This distribution, often traced back to \citet{Amoroso:1925} and later popularized by the works of \citet{Stacy:1962} and \citet{Stacy/Mihram:1965}, has been advocated for hydrological applications very early on \citep[see, e.g.,][]{Lienhard:1964, Lienhard/Meyer:1967}. One of its key advantages is its flexibility: it encompasses several well-known distributions as special cases, including the gamma and Weibull distributions, and approaches the lognormal distribution as a limiting case. In a particularly elegant contribution, \citet{Prentice:1974} showed how the lognormal distribution could be embedded as a proper special case within the generalized gamma family. Building on the same ideas, \citet{Perreault/Pesenti/Reid:2025+} proposes practical approximations that make it possible to implement this extension in a stable and reliable manner. In what follows, we adopt this extended version, referred to simply as the generalized gamma distribution.
When paired with the aforementioned seasonal modelling framework, this choice allows for a continuous and flexible representation of seasonal dynamics, accommodating changes in distributional shape throughout the year. As we highlight, it captures the empirical variability observed in river flow data quite well.

The paper is structured as follow. Section~\ref{sec:basic-assumptions} introduces the model for daily univariate environmental data, with a focus on purely seasonal model and seasonal models with time trends. Section~\ref{sec:model-fitting} is devoted to fit the misspecified model, model selection and analysis. In Section~\ref{sec:application}, we illustrate how to fit a seasonal model with time trends to real data from the Fraser River. We first analyse a single station and then illustrate how to jointly model multiple stations. Section~\ref{sec:conclusion} concludes and provides thoughts for future studies.

\section{Daily environmental model} \label{sec:basic-assumptions}

\subsection{Model setting}

To make the scope of our methodology explicit, we begin by providing general assumptions under which the proposed model is developed. Let $\mathcal{T} \subset \mathbb{R}$ be a set of $N$ ordered timestamps, possibly irregularly spaced, and $\bm{X} = (X_t)_{t \in \mathcal{T}}$ be some available data of interest. The core assumption, specifically designed for environmental data, is that $\bm{X}$ comes from a process $(X_t)_{t \in \mathbb{R}}$ that is seasonal in the following weak sense.
For all $t \in \mathbb{R}$, let $F_t$ be the cumulative distribution function (cdf) of $X_t$ and suppose that $F_t(\cdot) = F(\cdot|\bm{\theta}_t)$ for some parameter $\bm{\theta}_t$ that varies with the time index $t$; without loss of generality, assume that $\bm{\theta}_t = (\mu_t,\sigma_t,\nu_t)$.
To incorporate some seasonality and long-term variation in the parameters, we adopt what is essentially a generalized additive model for location, shape, and scale \citep[GAMLSS,][]{Rigby/Stasinopoulos:2005}, using covariates that depends exclusively on the time index $t$. Specifically, we let
\begin{equation} \label{eq:linear-predictor-general}
    g_{\mu}(\mu_t) = \bm{a}_{\mu t}^\top \bm{\beta}_{\mu}\;, \qquad g_{\sigma}(\sigma_t) = \bm{a}_{\sigma t}^\top \bm{\beta}_{\sigma}\quad \text{and} \quad g_{\nu}(\nu_t) = \bm{a}_{\nu t}^\top \bm{\beta}_{\nu}\;,
\end{equation}
where $(g_{\mu}, g_{\sigma}, g_{\nu})$ are known monotonic and differentiable link functions, $(\bm{a}_{\mu t}, \bm{a}_{\sigma t}, \bm{a}_{\nu t})$ are the time-based covariates, and $(\bm{\beta}_{\mu}, \bm{\beta}_{\sigma}, \bm{\beta}_{\nu})$ are the unknown coefficients to be estimated. 

In the original formulation of \cite{Perreault/Pesenti/Reid:2025+}, the model is purely seasonal in that it supposes $F_t = F_{t + p}$ for some period or cycle length $p$. In this paper we work with $p \approx 365.25$, which corresponds to the length of the solar year. Such a model is built by restricting oneself to periodic functions of $t$ when defining $(\bm{a}_{\mu t}, \bm{a}_{\sigma t}, \bm{a}_{\nu t})$; e.g., Fourier basis functions (i.e., pairs of cosine and sine functions) or cyclic B-splines. Here, we extend their framework by further including covariates defined via polynomials of $t$, thus capturing more general non-stationary behaviours. Moreover, we consider covariates constructed as interactions between seasonal and drift components, which allows the seasonal pattern to evolve as time passes.

While the results in \citet{Perreault/Pesenti/Reid:2025+} do not strictly speaking allow for drift terms, they can be adapted to include polynomial terms or different basis functions.
The proofs in \citet{Perreault/Pesenti/Reid:2025+} however may require more substantial modifications when incorporating drift terms based on basis functions such as B-splines, since the coefficient associated with a given B-spline depends only on a localized subset of the data. This is enough to justify the use of linear drift terms and Fourier basis functions, which we prioritize in this paper.

\subsection{Univariate model for river flows} \label{sec:univariate}

As the above assumptions suggest, one of the core components of our approach is the univariate distribution $F_t$, which is ultimately based on a fixed function $F(\cdot|\bm{\theta}_t)$ with time-varying parameters $\bm{\theta}_t$.
For modelling river flows, we propose to use an extension, due to \citet{Prentice:1974}, of the three-parameter generalized gamma distribution described in various forms in, e.g., \citet{Stacy:1962}, \citet{Stacy/Mihram:1965}; see also \citet[Section~8.7]{Johnson/Kotz/Balakrishnan:1994} for a more comprehensive treatment.
Many common distributions are special cases of this family, most notably the Weibull, gamma and lognormal distributions, three popular choices in hydrology; see, e.g., Table~1 in \citet{Stacy/Mihram:1965} for a list associated with the standard (non-extended) generalized gamma distribution.
As its absence from the latter list suggest, the lognormal distribution is not a proper member of the standard generalized gamma family; it is however a limit case of it, as we explain below.

Following the parametrization in \citet[Section~19.4.3]{Rigby/al:2019}, we say that a positive random variable has a generalized gamma distribution, denoted $\mathrm{G}\Gamma(\mu_t,\sigma_t,\nu_t)$, when its density has the form
\begin{equation} \label{eq:gg-density}
    f(x_t|\mu_t,\sigma_t,\nu_t) := 
    \begin{cases}
        \displaystyle\frac{|\nu_t| \xi_t^{\xi_t} z_t^{\xi_t} \exp(-\xi_t z_t)}{x_t\Gamma(\xi_t)} & \nu_t \neq 0\\[1em]
        \displaystyle\frac{\exp\{-\log^2(x_t/\mu_t)/(2\sigma_t^2)\}}{x_t\sigma_t\sqrt{2\pi}} & \nu_t = 0\;,
    \end{cases}
\end{equation}
where $z_t := (x_t/\mu_t)^{\nu_t}$, $\xi_t := (\sigma_t\nu_t)^{-2}$, $\mu_t> 0$, $\sigma_t > 0$, and where $\Gamma(\cdot)$ denotes the Gamma function.
For $\nu_t \neq 0$, $f$ is the standard generalized gamma density, while for $\nu_t = 0$ it is the lognormal density with parameters $\log(\mu_t)$ and $\sigma_t^2$.
Given the parameter constraints, we set $g_\mu(x) = g_\sigma(x) = \log(x)$ (log links), thereby ensuring that $\mu_t$ and $\sigma_t$ always stay positive during estimation, and further set $g_\nu(x) = x$ (the identity link).

Despite its appearance, $f$ is indeed continuous in $\nu_t$ everywhere, including at $\nu_t = 0$.
In other words, the lognormal density defined for $\nu_t=0$ is the limit of $f$ as $\nu_t \to 0$, irrespective of the side from which $\nu_t$ approaches zero.
By showing that the score and Hessian of (a transformation of) this density exist at $\nu_t=0$, with the latter being positive definite, \citet{Prentice:1974} demonstrated that the lognormal could be included in the family without violating the standard regularity conditions for maximum likelihood estimation.
Doing so does however introduce some practical challenges due to the numerical instability in the evaluation of $f$ when $|\nu_t|$ is positive but small.
These technicalities and solutions for them are discussed in Section~\ref{sec:objective} below and in more details in \citet{Perreault/Pesenti/Reid:2025+}.

One of the key advantages of the extended generalized gamma distribution, particularly in river flow modelling, is its flexible tail behaviour.
Unlike the gamma and lognormal distributions, not all moments of the generalized gamma distribution necessarily exist; indeed, their existence depends on the values of the parameters $\sigma_t$ and $\nu_t$, and more precisely $\sigma_t^2 \nu_t$, which we refer to as the tail-index coefficient.
For $X_t \sim \mathrm{G}\Gamma(\mu_t,\sigma_t,\nu_t)$ and $k \in \mathbb{N}_+$,
\begin{equation*} %\label{eq:moments}
    \mathbb{E}(X_t^k) = 
    \begin{cases}
        \displaystyle\mu_t^k \; \frac{\Gamma(\xi_t + k/\nu_t)}{\xi^{k/\nu_t} \Gamma(\xi)}  &\qquad \text{if } \sigma_t^2 \nu_t > -1/k \text{ and } \nu_t \neq 0\\[1em]
        \mu_t^k \, \exp(k^2 \sigma_t^2/2) &\qquad \text{if } \nu_t = 0\\[1em]
        \infty  &\qquad \text{if } \sigma_t^2 \nu_t \leqslant -1/k\;.
    \end{cases}    
\end{equation*}
Note that the $k$-th moment exists if and only if $\sigma_t^2 \nu_t > -1/k$. Thus, the larger the tail-index $\sigma_t^2 \nu_t$, the more moments exist, with all of them existing whenever $\nu_t \geqslant 0$.
Moreover, the mean and variance are 
\begin{subequations}\label{eq:moments}
    \begin{align} 
    \mathbb{E}(X_t) &=
    \begin{cases}
       \displaystyle
       \mu_t\, \frac{ \gamma_{t1}}{\xi^{1/\nu_t}\gamma_{t0}} & \quad \text{if } \nu_t \neq 0\\[1em]
        \mu_t \, \exp(\sigma_t^2/2) & \quad \text{if } \nu_t = 0\;,
        \quad \text{and} \quad
    \end{cases} 
    \\[2em]
    \mathbb{V}\mathrm{ar}(X_t) &= 
    \begin{cases}
       \displaystyle
       \mu_t^2\,  \frac{(\gamma_{t0}\gamma_{t2} -\gamma_{t1}^2)}{(\xi^{1/\nu_t}\gamma_{t0})^2} & \text{if } \nu_t \neq 0\\[1em]
        \mu_t^2 \, \{\exp(2 \sigma_t^2) - \exp(\sigma_t^2)\} & \text{if } \nu_t = 0\;,\\
    \end{cases} 
\end{align}
\end{subequations}
where $\gamma_{tk} := \Gamma(\xi_t + k/\nu_t)$, provided that $\sigma_t^2 \nu_t > -1$ and $\sigma_t^2 \nu_t > -1/2$, respectively.

\subsection{Purely seasonal model} \label{sec:linear-predictor-1}

To incorporate periodicity in the parameters $(\mu_t,\sigma_t,\nu_t)$, we include as covariates, in $(\bm{a}_{\mu t},\bm{a}_{\sigma t},\bm{a}_{\nu t})$ of \eqref{eq:linear-predictor-general}, so-called Fourier basis functions, that is, pairs of functions $C_k(t) = \cos(2\pi k t/p)$ and $S_k(t) = \sin(2\pi k t/p)$ for some integer values of frequencies $k$ and where $p\in \mathbb{N}_+$ is the cycle length.
Specifically, when defining a purely seasonal model, we suppose that for each parameter $\theta_t \in \{\mu_t, \sigma_t, \nu_t\}$ and all $t \in \mathbb{R}$, the link functions satisfy
\begin{equation} \label{eq:linear-predictor-1}
    g_{\theta}(\theta_t) := \bm{a}_{\theta t}^\intercal\bm{\beta}_\theta =
\beta_{\theta 0} + \sum_{k=1}^{d_\theta} \Big\{ \beta_{\theta k}^c C_k(t) + \beta_{\theta k}^s S_k(t) \Big\}\;,
\end{equation}
where $\bm{\beta}_\theta  := \big(\beta_{\theta 0}, \{\beta_{\theta k}^c, \beta_{\theta k}^s\}_{k \leqslant d_\theta}\big)$ concatenates $\beta_{\theta 0}$, $\beta_{\theta k}^c$, and $\beta_{\theta k}^s$ ($k \leqslant d_\theta$) to a vector of length $2d_\theta + 1$. Similarly we have a $2d_\theta + 1$ dimensional vector of covariates $\bm{a}_{\theta t}:=\big(1,  \{C_k(t) + S_k(t)\}_{k \leqslant d_\theta}\big)$ and $d_\theta$ is either specified by the user, or learned via model selection (see Section~\ref{sec:model-selection}).
Implicit in \eqref{eq:linear-predictor-1} is that we allow a given sine-cosine pair to be included only if all pairs of smaller integer frequency are also included.

To better interpret the linear predictor in \eqref{eq:linear-predictor-1}, note that for a given frequency $k$, the pair of coefficients $(\beta_{\theta k}^c,\beta_{\theta k}^s)$ can be jointly interpreted as a transformation of the amplitude and phase parameters $(\alpha_{\theta k},\rho_{\theta k})$ of a single frequency $k$ cosine term.
This is seen by letting $\beta_{\theta k}^c = \alpha_{\theta k} \cos(\rho_{\theta k})$ and $\beta_{\theta k}^s = \alpha_{\theta k} \sin(\rho_{\theta k})$, so that
$$
    \alpha_{\theta k} \cos\big(2\pi k t/p - \rho_{\theta k} \big) = \beta_{\theta k}^c C_k(t) + \beta_{\theta k}^s S_k(t)\;.
$$
While the amplitude-phase representation is more intuitive, the sine-cosine formulation enables expressing $g_\theta(\theta_t)$ as a linear predictor, which is better suited for optimization routines.

\subsection{Dynamic seasonal model} \label{sec:linear-predictor-2}

The model presented in Section~\ref{sec:linear-predictor-1} is essentially that proposed in \citet{Perreault/Pesenti/Reid:2025+}.
In the present paper, we further allow the periodic pattern, including the intercept, to evolve over time.
We do so by including the time-index $t$ as a covariate, effectively creating a linear time trend, as well as interactions between $t$ and the sine-cosine pairs.
In fact, as we mentioned in Section~\ref{sec:basic-assumptions}, the model can in theory also contain higher-order polynomials. However, including both higher-order polynomials and interactions with the sine-cosine terms may introduce unnecessary complexity, so we opt for a simpler model. Concretely, we consider 
\begin{equation} \label{eq:linear-predictor-2}
    g_{\theta}(\theta_t) = \beta_{\theta 0} + \beta_{\theta 0}^t  t  + 
    \sum_{k=1}^{d_\theta} \Big\{ \beta_{\theta k}^c C_k(t) + \beta_{\theta k}^s S_k(t) \Big\} +
    \sum_{k'=1}^{p_\theta} \Big\{ \beta_{\theta k'}^{ct} C_{k'}(t) + \beta_{\theta k'}^{st} S_{k'}(t) \Big\}\,  t\;,
\end{equation}
where $p_\theta$ is constrained such that $p_{\theta} \leqslant d_{\theta}$ and, like $d_\theta$, can be either specified by the user, or learned via model selection (see Section~\ref{sec:model-selection}). As before, interactions with sine-cosine pairs of a given frequency are permitted only if all lower-frequency pairs are also included, both individually and in interaction with $t$ (of frequency $k' = 1$). In particular, we require the inclusion of the linear time trend (i.e., the interaction at frequency zero) as a prerequisite for including any higher-frequency interactions. 

We summarize the structure of the linear predictor associated with $\theta_t$ using the notation $\mathcal{S}_\theta = (d_\theta,p_\theta)$, where $d_\theta,p_\theta \in \{0,1,\dots,\}$ and $d_\theta \leqslant p_\theta$. For example, $\mathcal{S}_\theta = (2,1)$ indicates a model with an intercept, two sine-cosine pairs (of frequency $k=1$ and $k=2$), a time trend, and an interaction between the frequency one pair and $t$. For purely seasonal models, we use the notation $\mathcal{S}_\theta = (d_\theta,-)$; e.g., $\mathcal{S}_\theta = (0,-)$ indicates a model with only an intercept.

\section{Model fitting procedure}\label{sec:model-fitting}
\subsection{Objective function} \label{sec:objective}

In Sections~\ref{sec:linear-predictor-1} and \ref{sec:linear-predictor-2}, we introduced coefficients $\bm{\beta} := (\bm{\beta}_\mu, \bm{\beta}_\sigma, \bm{\beta}_\nu)$ that, along with the link functions and design matrices, fully determine the value of the marginal parameters $(\mu_t,\sigma_t,\nu_t)$ at any time $t$.
Although our primary interest lies in the marginal parameters, model estimation is conducted by maximizing an objective that is a function of $\bm{\beta}$.
Specifically, given a dataset of river flows $\bm{X}$, we obtain estimates $\bm{\hat\beta}$ of $\bm{\beta}$ by maximizing
\begin{equation} \label{eq:objective}
\ell(\bm{\beta}|\bm{X}) := \sum_{t \in \mathcal{T}} \log f(X_t | \mu_t, \sigma_t, \nu_t)\;,    
\end{equation}
where $\bm{\mu} := (\mu_t)_{t \in \mathcal{T}} := g_{\mu}^{-1}(\bm{A}_{\mu}\bm{\beta}_{\mu})$, $\bm{\sigma} := (\sigma_t)_{t \in \mathcal{T}} := g_\sigma^{-1}(\bm{A}_{\sigma}\bm{\beta}_{\sigma})$ and $\bm{\nu} := (\nu_t)_{t \in \mathcal{T}} := g_\nu^{-1}(\bm{A}_{\nu}\bm{\beta}_{\nu})$.
Note that our objective function is in fact the log-likelihood for independent observations drawn from the marginal model of Section~\ref{sec:univariate}.
This simplifying (working) assumption of independence, that is, intentionally ignoring potential temporal dependence between observations, is made to avoid the complexities of modelling temporal structure, which is often challenging for natural phenomena such as river flows.

In view of this latter fact, and the GAMLSS-like structure of \eqref{eq:linear-predictor-2}, our model can be interpreted as a misspecified GAMLSS.
In cases where $\nu_t$ is bounded away from zero for all $t$, one may use existing statistical software that implements the standard generalized gamma distribution \citep[e.g., the \texttt{R} package \texttt{gamlss}][]{gamlss:2024, Rigby/al:2019} to maximize \eqref{eq:objective}.
However, we expect the lognormal distribution to provide a reasonable fit for river flow data, making it likely that $\nu_t$ hovers around zero for some $t$.
As a consequence, maximizing \eqref{eq:objective} as explicitly defined via \eqref{eq:gg-density} may be numerically unstable.
To alleviate these difficulties, we rely on the implementation of the extended generalized gamma distribution provided by \citet{Perreault/Pesenti/Reid:2025+}, which uses (Taylor and Stirling's formula) approximations to handle cases when $\nu$ is small or $\xi$ is large.

\subsection{Inference for misspecified models} \label{sec:inference}

Using independence as a working assumption simplifies the model but introduces complications for inference.
With correctly specified models, inference is usually based on the classical asymptotic theory that exploits the second Bartlett identity. The identity states that the Fisher information matrix $\mathcal{I}_{\bm{\beta}} := -\mathbb{E}\big(\nabla_{\bm{\beta}}^2 \ell(\bm{\beta}|\bm{X})\big)$ coincides with the variance of the score function $\mathcal{K}_{\bm{\beta}} := \mathbb{V}\mathrm{ar}\big(\nabla_{\bm{\beta}} \ell(\bm{\beta}|\bm{X})\big)$.
When the model is misspecified, this identity no longer holds.
Nevertheless, under suitable regularity conditions, the estimator $\bm{\hat\beta}$ remains asymptotically normal. 
Its covariance matrix takes the so-called sandwich form, that is, $\mathbb{V}\mathrm{ar}(\bm{\hat\beta}) \approx \mathcal{G}_{\bm{\beta}} := \mathcal{I}_{\bm{\beta}} \mathcal{K}_{\bm{\beta}}^{-1} \mathcal{I}_{\bm{\beta}}$. More precisely
\begin{equation} \label{eq:asymptotic}
    \sqrt{n}(\bm{\hat\beta} - \bm{\beta}) \rightsquigarrow \mathcal{N}(\bm{0}, \mathfrak{G}_{\bm{\beta}}^{-1})\quad \text{as } n := |\mathcal{T}| \to \infty\;.    
\end{equation}
where $\mathfrak{G}_{\bm{\beta}} := \mathfrak{I}_{\bm{\beta}} \mathfrak{K}_{\bm{\beta}}^{-1} \mathfrak{I}_{\bm{\beta}}$ and $\mathfrak{I}_{\bm{\beta}}$ and $\mathfrak{K}_{\bm{\beta}}$ are such that $\mathcal{I}_{\bm{\beta}}/n \to \mathfrak{I}_{\bm{\beta}}$ and $\mathcal{K}_{\bm{\beta}}/n \to \mathfrak{K}_{\bm{\beta}}$, as $n \to \infty$.

To perform inference based on \eqref{eq:asymptotic}, we require a consistent estimator $\hat{\mathcal{G}}_{\bm{\beta}}$ of $\mathcal{G}_{\bm{\beta}}$, in the sense that $\hat{\mathcal{G}}_{\bm{\beta}}/n \to \mathfrak{G}_{\bm{\beta}}$ in probability as $n\to\infty$, or alternatively, a resampling method that achieves this consistency implicitly.
In this work, we use a sandwich-type estimator, i.e., of the form $\hat{\mathcal{G}}_{\bm{\beta}}^{-1} = \hat{\mathcal{I}}_{\bm{\beta}}^{-1} \hat{\mathcal{K}}_{\bm{\beta}} \hat{\mathcal{I}}_{\bm{\beta}}^{-1}$, based on consistent estimators of $\mathcal{I}_{\bm{\beta}}$ and $\mathcal{K}_{\bm{\beta}}$ computed from the fitted null model.
The estimation of $\mathcal{I}_{\bm{\beta}}$ is relatively straightforward, but that of $\mathcal{K}_{\bm{\beta}}$ requires greater care due to the temporal dependence among the individual score contributions; as the usual assumption of independence among summands in $\nabla_{\bm{\beta}} \ell(\bm{\beta}|\bm{X}) = \sum_{t \in \mathcal{T}} \nabla_{\bm{\beta}} \ell(\bm{\beta}|X_t)$ does not hold. A common approach consists of using a weighted sample variance \citep[see, e.g.,][]{Andrews:1991, Lumley/Heagerty:1999}, where covariances between score terms are downweighted as their temporal separation increases. In this work, we adopt the Tukey-Hanning weighting scheme with a 31-day bandwidth, that is, assigning zero weight to covariances between observations separated by more than 31 days. See \citet{Perreault/Pesenti/Reid:2025+} for more details on the sandwich estimator that we use and some alternatives, and \citet{Lumley/Heagerty:1999} for a useful discussion that connects related approaches. 

Given an estimate of $\mathcal{G}_{\bm{\beta}}$, we can compute approximate confidence intervals for the parameters in the standard fashion using normal quantiles.
Similarly, hypothesis tests can be performed using test statistics of the form $(\bm{\hat\beta} - \bm{\beta})^\top \hat{\mathcal{G}}_{\bm{\beta}} (\bm{\hat\beta} - \bm{\beta})$, which in view of \eqref{eq:asymptotic} is asymptotically $\chi^2$ as $n\to\infty$.
For example, one may test the hypothesis $H_0 : \nu_t = 0$ for all $t$ (reducing the generalized gamma to a log-normal), which can be restated as $H_0 : \bm{\beta}_\nu = \bm{0}$, by comparing $\bm{\hat\beta}_{\nu}^\top \hat{\mathcal{G}}_{\bm{\beta},\nu} \bm{\hat\beta}_{\nu}$ to the quantiles of the $\chi_{d_\nu}^2$ distribution with $d_{\nu}$ degrees of freedom; here $\hat{\mathcal{G}}_{\bm{\beta},\nu}$ is the inverse of the submatrix of $\hat{\mathcal{G}}_{\bm{\beta}}^{-1}$ associated with $\bm{\hat\beta}_{\nu}$.

\subsection{Model selection} \label{sec:model-selection}

The construction of confidence intervals discussed in Section~\ref{sec:inference} assumes that a working model has been specified a priori, i.e., that for each parameter $\theta$, the quantities $d_{\theta}$ and $p_{\theta k}$ are given.
When the model structure is uncertain, a method is needed to select a suitable, yet parsimonious model.
This can be achieved either through penalization or by applying an appropriate information criterion.
Given the strict order in which we want to introduce new sine-cosine pairs, time trends and their interactions, we adopt a stepwise procedure that builds a sequence of candidate models of increasing complexity. 

To initialize the model structure, we use as linear predictors only intercepts:
\begin{equation} \label{eq:base-1}
\log(\mu_t) = \beta_{\mu0}\;, \quad \log(\sigma_t) = \beta_{\sigma0}\;, \quad \text{and} \quad \nu_t = \beta_{\nu0}\;,
\end{equation}
which, in the notation introduced in Section~\ref{sec:linear-predictor-2}, is summarized by $\mathcal{S}_\mu = \mathcal{S}_\sigma = \mathcal{S}_\nu = (0,-)$.
Then, given this model (or any subsequent one), we generate a set of candidate models (see below) and identify that corresponding to the highest relative increase of the objective function, given by
\begin{equation}\label{eq:relative-update}
    \frac{\ell_{k+1} - \ell_{k}}{q_{k+1} - q_k}\;,
\end{equation}
where $\ell_{k}$ and $\ell_{k+1}$ are the values of the objective (misspecified likelihood) for the base and candidate models, respectively, and $q_{k}$ and $q_{k+1}$ the total number of coefficients they respectively involve. Using the ratio of increases in the objective and number of parameters favours a fairer comparison of candidate models of distinct complexity.

When building a purely seasonal model, we generate the candidate models as follows. Given the model in \eqref{eq:base-1}, or any subsequent model, we generate six new candidates by further updating either of the three linear predictors in one of the following two ways: (i) $\mathcal{S}_\theta = (k,-) \leftarrow \mathcal{S}_\theta = (k+1,-)$, or (ii) $\mathcal{S}_\theta = (k,-) \leftarrow \mathcal{S}_\theta = (k+2,-)$. 
In other words, we increase the number of sine-cosine pairs by either one or two. Considering increases of size two allows us, using parallel computations, to speed up the exploration of the parameter space, but more importantly to overcome the possibility that a sine-cosine pair of low relative impact on the objective prevents the algorithm to reach a better model with a few more parameters; a phenomenon that we have witnessed in preliminary tests. We proceed similarly when building a dynamic seasonal model, by further allowing the update (iii) $\mathcal{S}_\theta = (k,r) \leftarrow \mathcal{S}_\theta = (k,r+1)$, and (iv) $\mathcal{S}_\theta = (k,r) \leftarrow \mathcal{S}_\theta = (k,r+2)$, provided that $r+1 \leqslant k$ and $r+2 \leqslant k$, respectively. When updating the model for first time, that is from $\mathcal{S}_\theta = (k,-)$ to  $\mathcal{S}_\theta = (k,c)$, $c \in \mathbb{N}$, we interpret the dash in $\mathcal{S}_\theta = (k,-)$ as $-1$. Note that when the base model is a purely seasonal, we do not consider the update (iv), which adds interactions between the time trend and the sin-cosine basis. This strategy introduces the time trend first before adding interaction terms. This leads to at most $12$ candidates models.

We iterate the above procedure multiple times to create a sequence of models $\{(\mathcal{S}_\mu^{(k)},\mathcal{S}_\sigma^{(k)},\mathcal{S}_\nu^{(k)})\}_{k=0}^K$, where $K$ is such that $(\mathcal{S}_\mu^{(K)},\mathcal{S}_\sigma^{(K)},\mathcal{S}_\nu^{(K)})$ is clearly overparametrized.
From this sequence, we select a model using Takeuchi's information criterion \citep[TIC][]{Takeuchi:1976}, defined as
\begin{equation} \label{eq:TIC}
    -\ell(\bm{\hat\beta}|\bm{X}) + 2\; \mathrm{trace}(\hat{\mathcal{K}}_{\bm{\beta}}\hat{\mathcal{I}}_{\bm{\beta}}^{-1})\;,
\end{equation}
where $\hat{\mathcal{I}}_{\bm{\beta}}$ and $\hat{\mathcal{K}}_{\bm{\beta}}$ are as described in Section~\ref{sec:inference} and are computed for each candidate model.
The second term in \eqref{eq:TIC} serves a role analogous to the dimensionality penalty in the more standard Akaike information criterion \citep[AIC][]{Akaike:1974}, but it is adapted to accommodate model misspecification. While it is common to select a final model by minimized TIC, we use it more as an analytic tool to guide model selection, to guard against the potential effects of the uncertainty in the estimation of $\mathcal{I}_{\bm{\beta}}$ and $\mathcal{K}_{\bm{\beta}}$.

\subsection{Diagnostics} \label{sec:qq-plots}

To assess the fit of a given model we analyse the normalized pseudo-observations (z-scores) given by $\Phi^{-1}(F(X_t|\mu_t,\sigma_t,\nu_t))$, where $F$ is the generalized gamma cdf and $\Phi^{-1}$ denotes the quantile function of the standard Gaussian distribution.
Expressing the residuals on the standard normal scale allows for intuitive assessment using standard normal-based diagnostics in statistical modelling.
In particular, we visually assess the fit by means of quantile-to-quantile plots (QQ-plots), that is, by plotting the z-scores against the theoretical quantile of the standard Gaussian distribution.
Since model performance may vary over the seasonal cycle, it is also useful to stratify the z-scores by month, to evaluate whether the model performs consistently throughout the year. Examples of this methodology are shown in Figures~\ref{fig:qq-A}--\ref{fig:qq-BC}.

\section{Application to river flow} \label{sec:application}

\subsection{Data and models}
\begin{figure}[t]
    \centering
    \includegraphics[width=.4\textwidth]{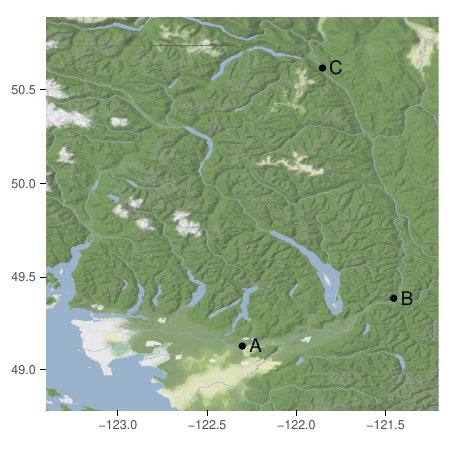}
    \hspace{.5cm}
    \includegraphics[width=.533\textwidth]{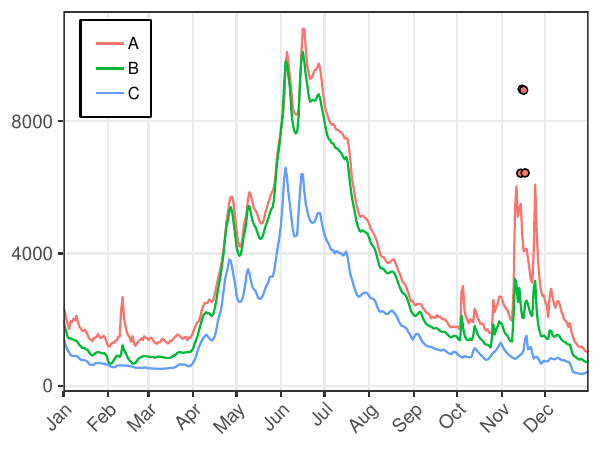}
    \caption{Left panel: Location of the three hydrometric stations on the Fraser River from which the data in Section~\ref{sec:application} were collected. Right panel: River flow series for the three stations in 1990, with four additional points marking the major flood event of November 2021 at station~A.}
    \label{fig:map-data}
\end{figure}

To illustrate our methodology, we consider river flow data (daily average flow in \SI{}{m^3/s}) from three distinct hydrometric stations located on the Fraser River in British Columbia (Canada), referred to as stations~A, B, and C; their locations are given by Figure~\ref{fig:map-data}.
As mentioned in the introduction, the Fraser river originates in the Canadian Rocky Mountains, passes through the Fraser Canyon and the Fraser Valley, and finally empties into the Strait of Georgia close to Vancouver. Station A lies at the end of the Fraser Valley and is known for flooding that incur large losses due to its proximity to the city of Vancouver. Station B is upstream and in the Fraser Canyon, thus already significantly closer to the mountains. Station C is in the Fraser Canyon is even further upstream, thus subject to significant snow coverage in winter.

The data consist of uninterrupted series from 1965-05-01 to 1992-12-31 for station~A ($\sim$27.67 years, 10,107 observations), 1912-03-01 to 2023-12-31 for station~B ($\sim$111.8 years, 40,848 observations), and 1951-08-12 to 2023-12-31 for station~C ($\sim$72.4 years, 26,440 observations).

We define three situations based on the available data, chosen to illustrate different aspects of the modelling framework.
We begin with station~A, which has the shortest series, and fit a purely seasonal model (Section~\ref{sec:linear-predictor-1}), as this is the setting where such a specification is most reasonable.
An additional 1,360 observations are available for this station, but these are mostly concentrated in the May–July months of 2000–2023, and addressing this imbalance lies outside the scope of the paper.
We nevertheless investigate the November 2021 flood shown in Figure~\ref{fig:map-data}, to illustrate the potential pitfalls of extrapolating beyond the data range by examining its associated estimated return period.
We then turn to station~B, which has the longest time data series, and fit the dynamic seasonal model (Section~\ref{sec:linear-predictor-2}). Finally, we illustrate how data from multiple stations, here stations B and C, can be combined to inform the choice of model structure, again in the context of a dynamic seasonal model.

As explained in Section~\ref{sec:model-selection}, we use TIC to guide us with the selection of the final model. For each of the applications, we include in the computation of $\mathcal{K}_{\bm{\beta}}$ all covariance terms corresponding to pairs of observations separated by up to 31 days, which we weight using the Tukey-Hanning scheme briefly described in Section~\ref{sec:inference}; see also \citet{Andrews:1991}. For the last application, which involves two stations, we also include the between-station covariances that fall within the considered time-windows.
The 31-day bandwidth parameter corresponds roughly to $n^{3/8}$, $n^{1/3}$, and $n^{4/9}$ for the three applications in their respective order. Some tests with smaller (20) and larger (60) bandwidth parameters suggests that the results are robust to this choice.

\subsection{Purely seasonal model}\label{sec:application-purely}

We begin by fitting a purely seasonal model to the $n=10,107$ observations from station~A. To provide a more concrete description of our procedure, we explain the first few steps of the algorithm. 
\begin{figure}[t]
    \centering
    \includegraphics[width=.9\textwidth]{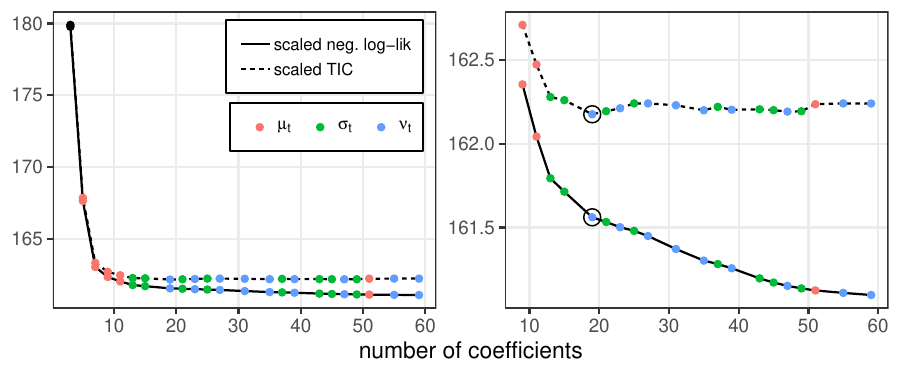} 
    \caption{Scaled negative log-likelihood (solid line) and scaled Takeuchi Information Criterion (dashed line) for the purely seasonal model fitted to station~A. Values are plotted against the total number of coefficients used in the linear predictors for $\mu_t$, $\sigma_t$, and $\nu_t$; the scaling factor for the y-axis is $10^{-3}$. Left panel: Overall behaviour across all models on the path. Right panel: Closer view of the region where the criteria stabilize, with the selected model circled.}
    \label{fig:tic-A}
\end{figure}
A depiction of the full model selection procedure is given by Figure~\ref{fig:tic-A}. It displays the evolution of the TIC criterion, the parameter whose linear predictor was updated at each step, and the total number of coefficients after each update.
As shown therein, fitting the base model in \eqref{eq:base-1} results in a TIC value of $\sim$179,893.
From this initial model, the algorithm evaluates potential updates to the linear predictor of each parameter $\theta_t \in \{\mu_t, \sigma_t, \nu_t\}$. That is for a given parameter, we consider the two candidate linear predictors given by
$$
    g_{\theta}(\theta_t) = \beta_{\theta0} + \beta_{\theta1}^c C_1(t) + \beta_{\theta1}^s S_1(t)\;, \quad \text{and} \quad g_{\theta}(\theta_t) = \beta_{\theta0} + \sum_{k=1}^2 \beta_{\theta k}^c C_k(t) + \beta_{\theta k}^s S_k(t)\;,
$$
where $g_\theta$ is the parameter-dependent link function. The two models are denoted $\mathcal{S}_\theta=(1,-)$ and $\mathcal{S}_\theta=(2,-)$, respectively.
Among the six models generated (two for each of $\{\mu_t, \sigma_t,\nu_t\}$), $\mathcal{S}_\mu(1,-)$ achieves the largest decrease in relative negative log-likelihood per additional parameter (see criterion \eqref{eq:relative-update}), and is therefore selected as the new base model.
In fact, this update, going from $\mathcal{S}_\mu=(k,-)$ to $\mathcal{S}_\mu=(k+1,-)$, is applied four times in a row in the first steps of the algorithm, resulting in the intermediate model given by $\mathcal{S}_\mu=(4,-)$ and $\mathcal{S}_\sigma = \mathcal{S}_\nu=(0,-)$, and a cumulative reduction of $\sim$17,421 in TIC.
The first two updates account for most of this reduction, $\sim$12,033  and $\sim$4,526, respectively.
This is intuitive, as $\mu_t$ is closely linked to the mean of the distribution and thus captures the strong seasonal pattern evident in the right panel of Figure~\ref{fig:map-data}.

Subsequently, the algorithm applies two analogous updates to $\sigma_t$ yielding $\mathcal{S}_\sigma=(2,-)$, and a single update to $\nu_t$ yielding $\mathcal{S}_\nu=(2,-)$. This model, given by
$\mathcal{S}_\mu=(4,-)$ and $\mathcal{S}_\sigma = \mathcal{S}_\nu=(2,-)$,
corresponds to a TIC value of $\sim$162,175 and turns out to be the minimum attained by any of the 23 models visited by the algorithm before the early stopping criterion (15 iterations without improving the current best TIC) was triggered. 
Note that, after the $8$th iteration, some updates do lead to decreases in TIC relative to the immediately preceding model, but these reductions are not large enough to offset earlier increases.
We conclude that the 8th model is the most appropriate choice for the final model.

\begin{figure}[p]
    \centering
    \includegraphics[width=.4\textwidth]{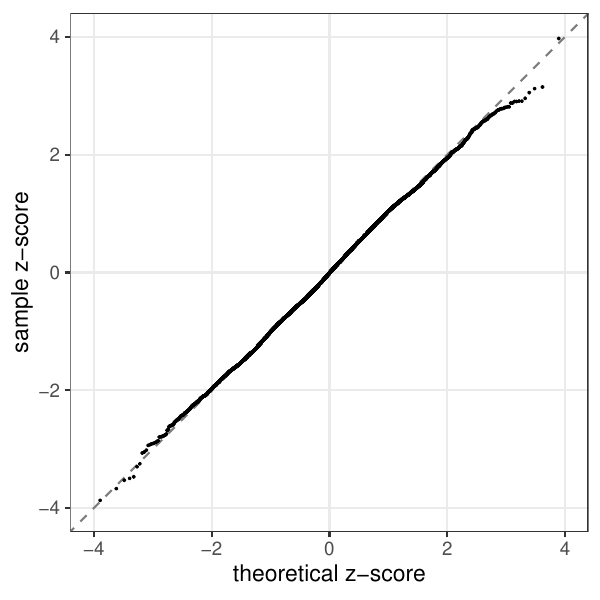}
    \includegraphics[width=.5\textwidth]{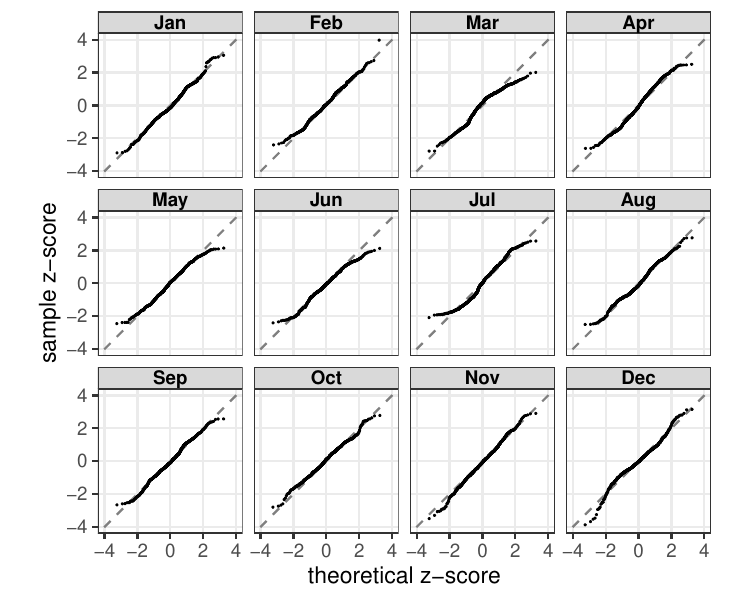}
    \caption{QQ-plots of normalized residuals for the purely seasonal model fitted to station~A. Left panel: Global QQ-plot. Right panel: Month-specific QQ-plots.}
    \label{fig:qq-A}
\end{figure}

We assess the fit provided by our final model using the QQ-plots described in Section~\ref{sec:qq-plots} and plotted in Figure~\ref{fig:qq-A}. The QQ-plot based on the full dataset (left panel of Figure~\ref{fig:qq-A}) suggests that the model performs very well overall. However, the QQ-plots specific to the months of March--July (see right panel of Figure~\ref{fig:qq-A}) show some noticeable deviations in the tails. These may reflect the absence of a time trend in the model, which could disproportionately affect those months. Alternatively, the data may not be rich enough to support the addition of higher-frequency terms, even if such features would ultimately be necessary to capture the seasonal variation for these months. In this context, using B-splines might offer a more flexible alternative. It is important to note, however, that departures from a straight line can be more pronounced than usual when the residuals are not independent, as is the case here, and should thus be interpreted with caution.

\begin{figure}[p]
    \centering
    \includegraphics[width=.9\textwidth]{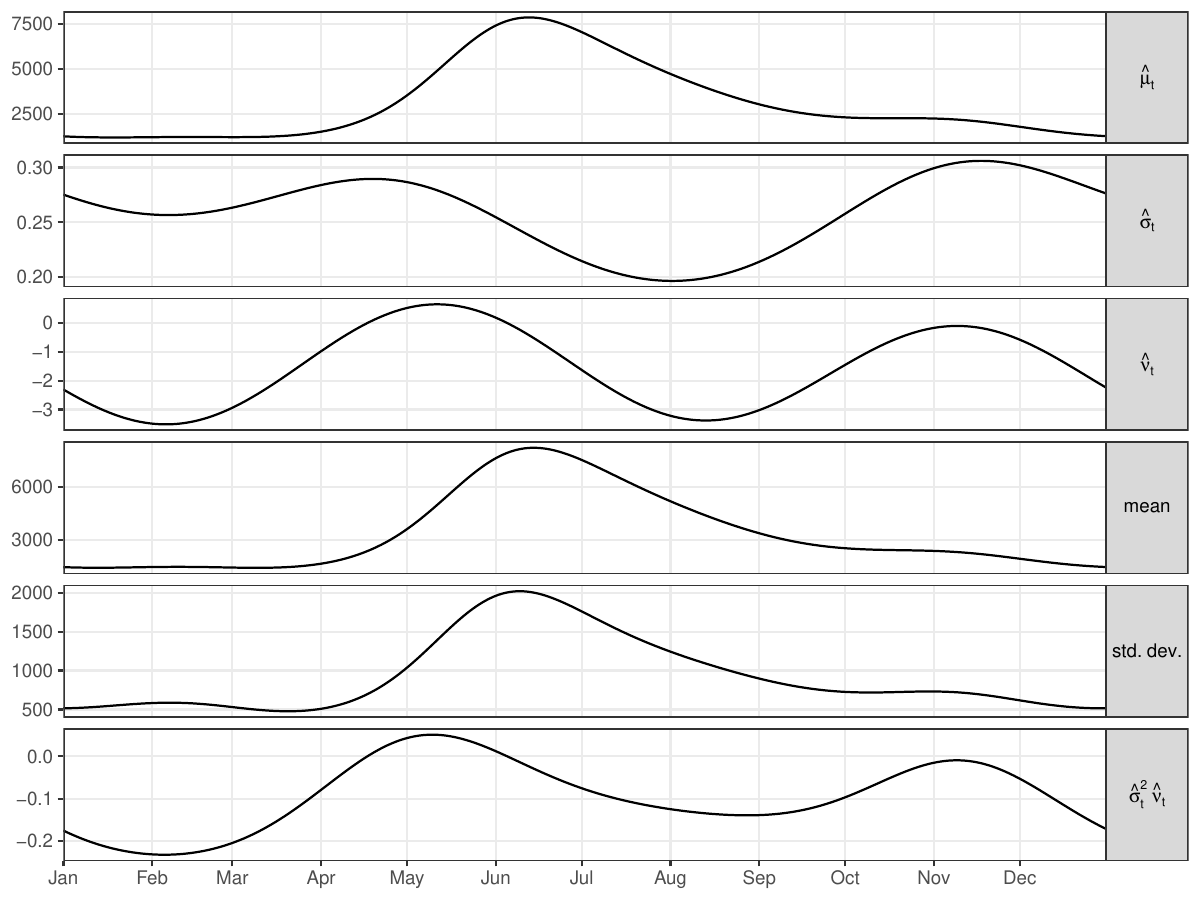}
    \caption{Estimated generalized gamma parameters $(\mu_t, \sigma_t, \nu_t)$ and associated statistics (mean, standard deviation, and tail-index $\sigma_t^2 \nu_t$) for the purely seasonal model fitted to station~A, as a function of the time of the year.}
    \label{fig:param-A}
\end{figure}

The first three panels of Figure~\ref{fig:param-A} depict the seasonal evolution of the estimated parameters, the mean and standard deviation (given in \eqref{eq:moments}), and $\hat\sigma_t^2 \hat\nu_t$, which can be interpreted as a tail-index. As mentioned in Section \ref{sec:univariate}, the tail-index determines which moments of $X_t$ exists, indeed $\hat\sigma_t^2 \hat\nu_t> -1/k$ means that all moments up to order $k$ are finite. 
In accordance with the relatively simple form of the linear predictors, the estimated parameters vary smoothly over the year. Recall that we fit a purely seasonal model, thus, the parameters vary periodically and are the same for different years.
For the location parameter $\hat\mu_t$, we observe in Figure~\ref{fig:param-A} a relatively sharp increase during spring, followed by a gradual decrease and a stabilization through the summer and fall, consistent with seasonal streamflow dynamics.
This is also reflected in the shape of the mean and standard deviation, which are driven mostly by the location parameter.
The patterns in $\hat\sigma_t$ and $\hat\nu_t$ are similar to one another, but they attain their peak (in absolute value) at different times. Their combined effect creates a noticeable bump in the standard deviation during the later part of the winter (February) and, to a lesser extent, in the later part of fall (November). 
The bottom panel displaying $\hat\sigma_t^2 \hat\nu_t$ suggests that the first four moments exist throughout the seasonal cycle, as $\hat\sigma_t^2 \hat\nu_t \geqslant -1/4$ for all $t$. In contrast, the fifth moment appears to diverge during the later part of the winter. Although initially surprising, a tentative explanation is that snow cover may be substantial at this time of year (note that the Fraser River originates in the Rocky Mountains), which could in turn allow for unusually large increases in river flow.

While these results are sensible, it is important to keep in mind that the cyclostationarity assumption might be too strong even for such a relatively short time series. This manifest itself, for example, in the estimated return period of the largest flow recorded during the November 2021 flood (on the 15th, see Figure~\ref{fig:map-data}), which is well outside the range of the training data.
This model yields a return period of 411,968 years, a highly implause result, even considering that the return period is day-specific (i.e., we expect the river flow \textit{on November 15th} to exceed that number once every 411,968 years). This result, and for that matter a careful analysis of the data itself, point to the presence of a systematic time trend. Since the flood lies far outside the range of the training sample, it is unsurprising that the purely seasonal model fails to provide a meaningful estimate. When we extend the model to allow linear time trends (without interactions), the estimated day-specific return period decreases to about 129 years, which is of a far more realistic magnitude.

\subsection{Dynamic seasonal model} \label{sec:application-dynamic}

\begin{figure}[t]
    \centering
    \includegraphics[width=.9\textwidth]{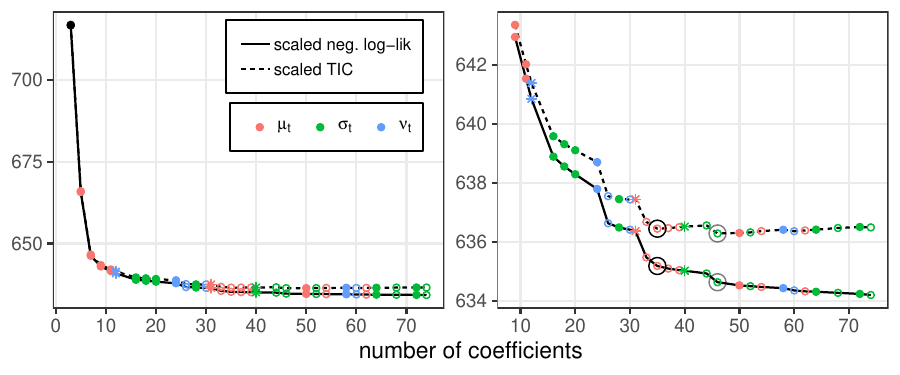} 
    \caption{Scaled negative log-likelihood (solid line) and scaled Takeuchi Information Criterion (dashed line) for the dynamic seasonal model fitted to station~B. Values are plotted against the total number of coefficients used in the linear predictors for $\mu_t$, $\sigma_t$, and $\nu_t$; the scaling factor for the y-axis is $10^{-3}$. Left panel: Overall behaviour across all models on the path. Right panel: Closer view of the region where the criteria stabilize, with the selected model circled in black and the TIC-optimal model circled in gray.}
    \label{fig:tic-B}
\end{figure}

We next fit a seasonal model with time trends and interactions to the $n=40,848$ observations from station~B. A depiction of the model path travelled by the algorithm is given in Figure~\ref{fig:tic-B}.
It shows that, the first coefficients added are a few sine-cosine pairs for $\mu_t$, as in the previous case. A few sine-cosine pairs are also again included for $\sigma_t$ among the first steps, but this time they are preceded by a time trend for $\nu_t$.
After these, two sine-cosine pairs are added for $\nu_t$ and the remaining steps involve mostly time trends or interactions. The algorithm was stopped after 10 iterations that did not improve the current best model in terms of TIC.
As expected, the TIC continues to decrease further along the model selection path, suggesting that models with more parameters remain justifiable -- which is in contrast to the more limited dataset from station~A -- resulting in a TIC-optimal model with significantly more coefficients, 46 in total. This model, given by $\mathcal{S}_\mu=(4,4)$, $\mathcal{S}_\sigma = (5,4)$ and $\mathcal{S}_\nu=(2,2)$ (circled in gray in Figure~\ref{fig:tic-B}), does not, however, seem significantly more advantageous than the model obtained five iterations earlier, which contains 35 coefficients (circled in black in Figure~\ref{fig:tic-B}). Upon further inspection, we chose this latter more parsimonious model, given by $\mathcal{S}_\mu=(4,2)$, $\mathcal{S}_\sigma = (5,-)$ and $\mathcal{S}_\nu=(2,2)$.

\begin{figure}[p]
    \centering
    \includegraphics[width=.4\textwidth]{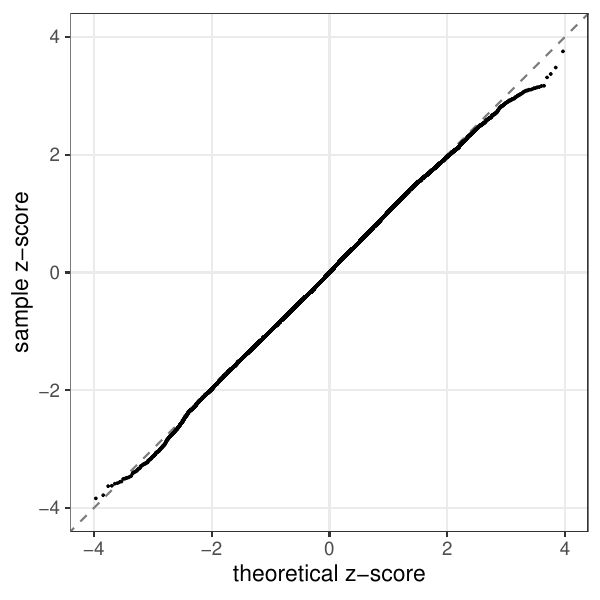}
    \includegraphics[width=.5\textwidth]{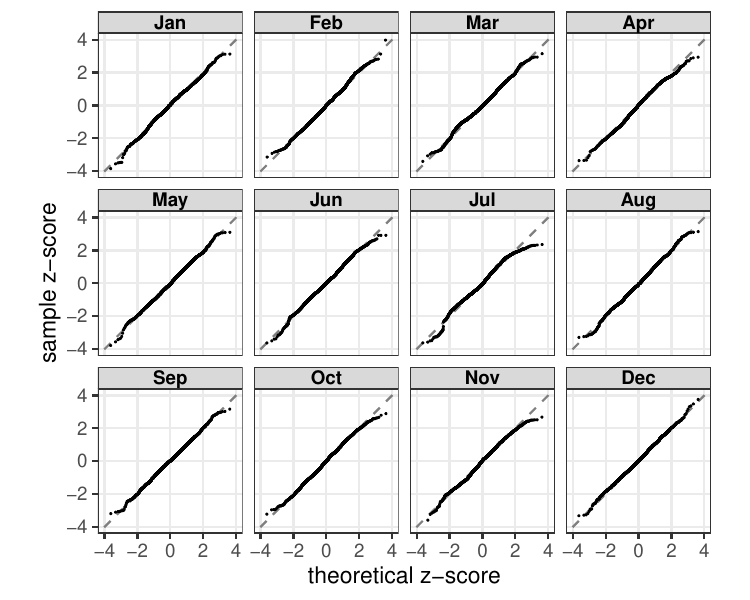}
    \caption{QQ-plots of normalized residuals for the dynamic seasonal model fitted to station~B. Left panel: Global QQ-plot. Right panel: Month-specific QQ-plots.}
    \label{fig:qq-B}
\end{figure}

We again assess the fit using the QQ-plots described in Section~\ref{sec:qq-plots}; these are shown in Figure~\ref{fig:qq-B}. As with the purely seasonal model selected for station~A in Section~\ref{sec:application-purely}, the QQ-plot based on the full dataset suggest that the fit is overall very good. In contrast to the model for station~A, the month-specific QQ-plots, especially those associated with the months of March--July, are significantly more satisfying. We find the fit to be convincing, indicating that the extended generalized gamma distribution is well suited for modelling river flow data.

\begin{figure}[p]
    \centering
    \includegraphics[width=.9\textwidth]{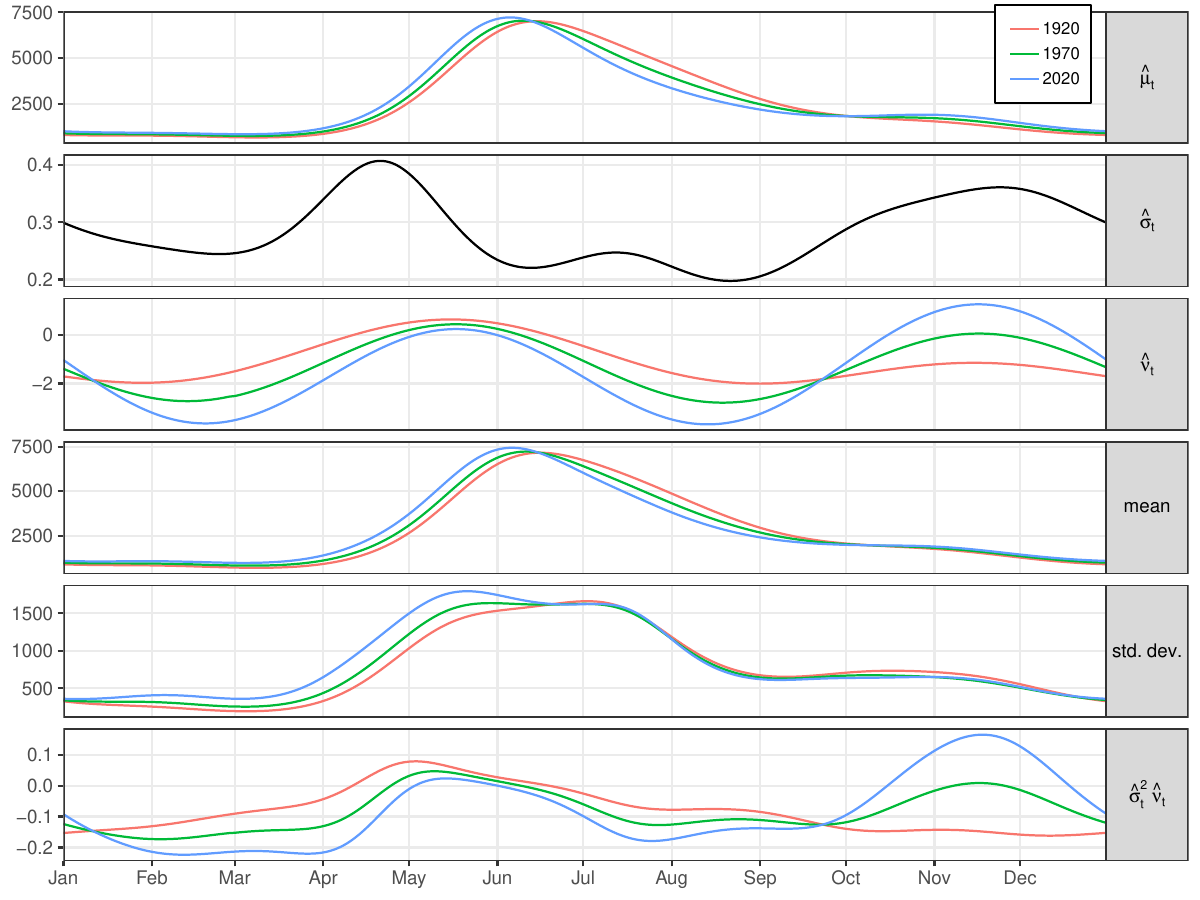}
    \caption{Estimated generalized gamma parameters $(\mu_t, \sigma_t, \nu_t)$ and associated statistics (mean, standard deviation, and tail-index $\sigma_t^2 \nu_t$) for the dynamic seasonal model fitted to station~B, as a function of the date for years 1920, 1970, and 2020.}
    \label{fig:param-B}
\end{figure}

To visualize the estimated parameters and related quantities, it is not sufficient, due to the presence of long term time trends, to report a single seasonal cycle (as in Figure~\ref{fig:param-A}).
To account for the variations over the years, we report in Figure~\ref{fig:param-B} the parameter values as well as the corresponding mean, standard deviation and tail-index, for three representative years (1920, 1970, and 2020), spanning most of the data range. We observe the same broad seasonal patterns as for station~A, and thus focus here on the aspects linked to the inclusion of time trends and interactions.
One noticeable feature is that the spring peak of the estimated location parameter $\hat{\mu}_t$, as well as the estimated mean, occurs progressively earlier over time.
While we do not claim expertise on the subject, this pattern seems consistent with expectations related to climate change.
Another feature is the absence of long-term trend in the final model for $\sigma_t$. 
This, however, should be interpreted with caution, since the standard deviation, higher-moments and tail-index, which all involve other parameters, still exhibit long-term trends.
In particular, the results suggest that the tail of the distribution becomes heavier over time during late winter and early spring, and lighter during autumn. It is worth noting, however, that the estimated fourth moment remains finite throughout the entire period.
More broadly, these results are consistent with popular criticisms about the assumption of stationarity (cyclostationarity) when dealing with environmental time series. This is especially true for the return period estimated in Section~\ref{sec:application-purely}, which corresponds to a date well outside the training data.

\subsection{Joint dynamic seasonal models}

In practice, combining information across nearby stations can provide more stable estimates, though there is a risk of imposing constraints that may not fully capture local differences. While directly sharing parameter values across stations would be inappropriate, sharing structural components provides a more plausible compromise. While the true structure for each station likely differs, pooling the data could reduce the variance enough to outweigh the potential bias introduced by this constraint. To examine this, we jointly fit the dynamic seasonal models to data from stations B and C, recall that stations B and C are on the Fraser River and that station B is downstream of station C (see Figure \ref{fig:map-data}). For the fit, we use the 26,440 dates (1951--2023) for which both stations have observations. More specifically, we fit a distinct model to each station, with parameter values estimated independently, while imposing a shared structure through identical $\mathcal{S}_\mu$, $\mathcal{S}_\sigma$, and $\mathcal{S}_\nu$. The  model selection procedure is as in Section \ref{sec:application-dynamic} with the difference that the objective is the sum of the objective functions of each station-specific model.

\begin{figure}[t]
    \centering
    \includegraphics[width=.9\textwidth]{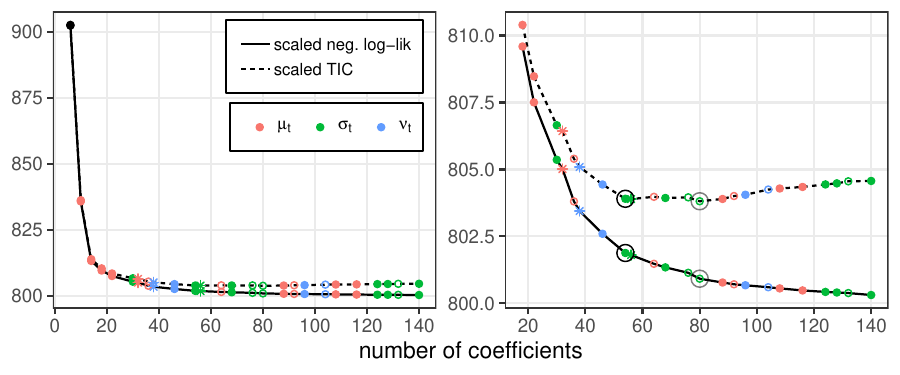} 
    \caption{Scaled negative log-likelihood (solid line) and scaled Takeuchi Information Criterion (dashed line) for the joint dynamic seasonal model fitted to stations~B and C. Values are plotted against the total number of coefficients used in the linear predictors for $\mu_t$, $\sigma_t$, and $\nu_t$; the scaling factor for the y-axis is $10^{-3}$. Left panel: Overall behaviour across all models on the path. Right panel: Closer view of the region where the criteria stabilize, with the selected model circled in black and the TIC-optimal model circled in gray.}
    \label{fig:tic-BC}
\end{figure}

\begin{figure}[p]
    \centering
    \includegraphics[width=.4\textwidth]{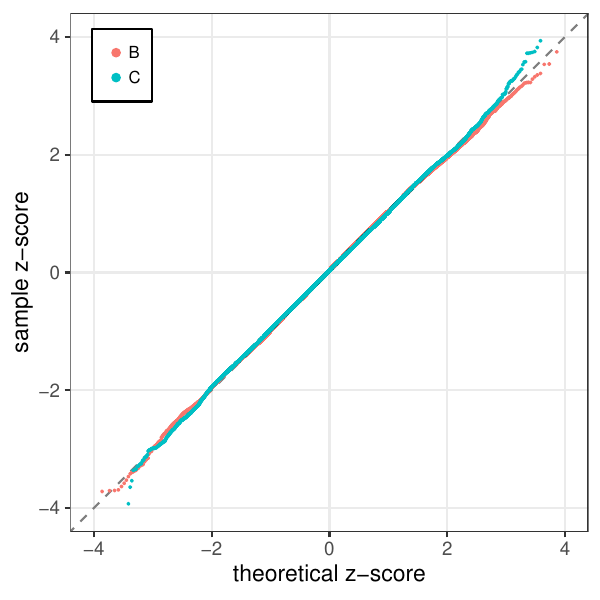}
    \includegraphics[width=.5\textwidth]{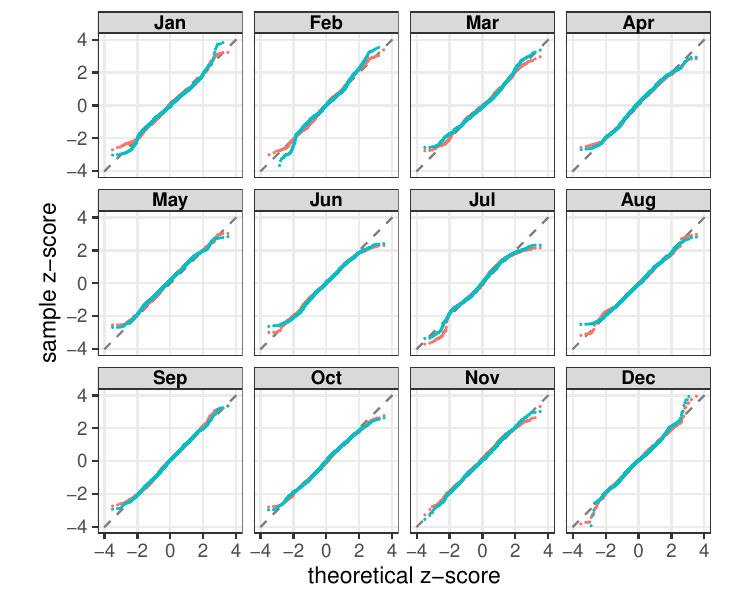}
    \caption{QQ-plots of normalized residuals for the joint dynamic seasonal model fitted to station~B and C. Left panel: Global QQ-plot. Right panel: Month-specific QQ-plots.}
    \label{fig:qq-BC}
\end{figure}

The path travelled by the algorithm is reported in Figure~\ref{fig:tic-BC}.
We obserce a similar pattern as for the dynamic model for station~B alone in Section~\ref{sec:application-dynamic}, and again we favour a simpler model, namely that given by $\mathcal{S}_\mu=(4,1)$, $\mathcal{S}_\sigma = (4,-)$ and $\mathcal{S}_\nu=(2,0)$ (circled in black in Figure~\ref{fig:tic-BC}) to the TIC-optimal one, given by $\mathcal{S}_\mu=(4,3)$, $\mathcal{S}_\sigma = (4,3)$ and $\mathcal{S}_\nu=(2,0)$ (circled in gray in Figure~\ref{fig:tic-BC}). Comparing the final model to that for station B in Section~\ref{sec:application-dynamic}, we obtain here a model with fewer time trend coefficients, which is consistent with the fact that the time series under study here covers a significantly shorter window of time; these results also carries over to the comparison of the two TIC-optimal models. The QQ-plot assessment for the two stations, shown in Figure~\ref{fig:qq-BC}, is broadly similar to that observed in Section~\ref{sec:application-dynamic}.
\begin{figure}[p]
    \centering
    \includegraphics[width=.9\textwidth]{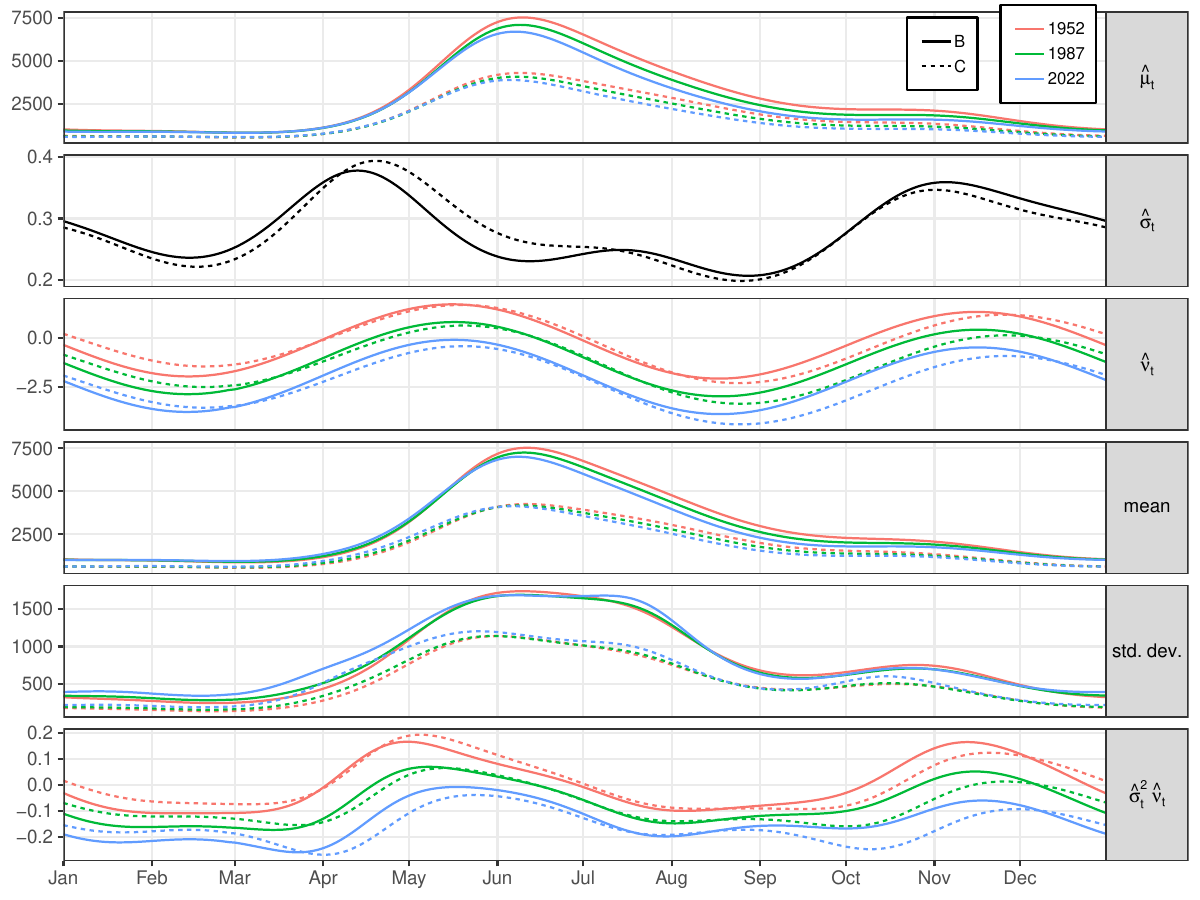}
    \caption{Estimated generalized gamma parameters $(\mu_t, \sigma_t, \nu_t)$ and associated statistics (mean, standard deviation, and tail-index $\sigma_t^2 \nu_t$) for the joint dynamic seasonal model fitted to stations~B and~C, as a function of the date for years 1952, 1987, and 2022.}
    \label{fig:param-BC}
\end{figure}

Similar to the dynamic model for station B of Section~\ref{sec:application-dynamic}, Figure~\ref{fig:param-BC} presents the estimated parameters and related quantities for three representative years spanning the range of the data: 1952, 1987, and 2022. This time, however, the figure now depicts two sets of parameters, one for each station. 
We begin by comparing the present results for station B with those obtained in Section~\ref{sec:application-dynamic}. Recall that for the single-station model for B, we used all available data for that station, i.e., from 1912 to 2023, while for the joint-station model, that is fitting stations B and C jointly, we only consider data from 1951 to 2023, the dates when both stations have data available.
The two models are broadly consistent, with the main discrepancy arising in $\hat{\nu}_t$, which exhibits stronger temporal variation (through interaction effects) in the single-station model. This pattern suggests that the earlier data (pre-1950) exert substantial influence on the fit, indicating that a more deliberate treatment of long-term time trends may be warranted. In particular, the pronounced rise in $\hat{\nu}_t$ during more recent winters in the single-station model appears questionable, as it may reflect an artifact introduced during optimization to better accommodate early observations.

When comparing the two stations, the main difference emerges in $\hat\mu_t$. We observe that station C has overall a smaller amount of river flow, which is consistent with the fact that B is downstream from station C and the watershed of station B is significantly larger than that of C.
By contrast, the estimates of $\sigma_t$ and $\nu_t$ are very similar across stations, raising the question of whether additional constraints in the specifications of their linear predictors should be considered; we return to this matter in the discussion.
These patterns extend to the corresponding mean, standard deviation, and tail-index quantities. The mean and standard deviation for each station exhibit broadly similar shapes, with differences that naturally mirror those seen in $\hat\mu_t$, while the estimated tail-indices are closely aligned across the two stations.

\section{Discussion}\label{sec:conclusion}

We offer a convenient and intuitive framework for studying environmental processes, leveraging the theory of misspecified models to avoid explicitly specifying temporal dependence. Conceptually, the model is a distributional regression based on periodic and smoothly varying functions of the time index. In this sense, it can be interpreted as a dynamic climatology: dynamic both because it is conditional on the time of year and because it accounts for changes in climate over time. Indices representing large-scale oscillatory phenomena (e.g., the North Atlantic Oscillation or El Niño–Southern Oscillation indices) could be incorporated to account for quasi-periodic variations in climate, if such conditions are desired \citep[see, e.g.,][]{Machado/al:2015}.

Within our proposed framework, we find that the extended generalized gamma distribution provides a particularly useful family for modelling daily river flows. Beyond offering a flexible and generally good fit, it also allows us to indirectly assess which of two widely used distributions in hydrology, the log-normal ($\nu_t = 0$) and the gamma ($\nu_t = 1$), is more appropriate in a given setting. Our empirical results suggest that, at least for the stations considered here, the log-normal distribution generally provides a better fit than the gamma. At the same time, there are indications that the underlying process may shift between regimes during the seasonal cycle, at times resembling a gamma distribution, at others a log-normal, and in some periods extending beyond either classical form, generally on the log-normal side.

For the sake of simplicity, we have not accounted for several potentially important issues. For instance, changes in the observational record could arise from instrument replacement, site relocation, or other measurement artifacts, rather than genuine hydrological variation. For this reason, any applied use of the methodology should ideally be undertaken in collaboration with domain experts (e.g., hydrologists in the case of river flow modelling) familiar with the data sources and their limitations.

Finally, our applications underscore certain technical considerations that merit further attention. The use of Fourier basis functions, while convenient, imposes global smoothness and can generate artifacts that might be avoided with alternative representations such as cyclic B-splines, which act locally. Our model does allow for structural changes through time, but very long data series may require more flexible specifications for the trend component; such flexibility, however, comes with the risk of overfitting rather than underfitting. Other directions worth investigating include addressing seasonal imbalance through appropriate weighting schemes to correct distortions from uneven data distribution across the seasonal cycle, and enhancing the model’s ability to capture tail behaviours, by potentially directly modelling extreme events using extreme-value distributions.

\vspace{2em}
\noindent
\textbf{Acknowledgments}
S. Pesenti would like to acknowledge support from the Natural Sciences and Engineering Research Council of Canada (RGPIN-2025-05847) and from the Canadian Statistical Sciences Institute (CANSSI). S. Perreault and S. Pesenti are grateful for the support from the Data Science Institute.

\printbibliography

\end{document}